\documentclass[aps,twocolumn,showpacs,preprintnumbers,amsmath,amssymb]{revtex4}
\usepackage{colordvi}
\usepackage{amssymb}
\usepackage{amsmath}
\usepackage{epsf}
\usepackage{mathrsfs}
\usepackage{graphicx}
\usepackage{appendix}
\begin{document}
\def \beq{\begin{equation}}
\def \eeq{\end{equation}}
\def \bea{\begin{eqnarray}}
\def \eea{\end{eqnarray}}
\def \bem{\begin{displaymath}}
\def \eem{\end{displaymath}}
\def \P{\Psi}
\def \Pd{|\Psi(\boldsymbol{r})|}
\def \Pds{|\Psi^{\ast}(\boldsymbol{r})|}
\def \Po{\overline{\Psi}}
\def \bs{\boldsymbol}
\def \bl{\bar{\boldsymbol{l}}}
\title{  Density Wave -Supersolid and Mott Insulator-Superfluid transition in presence of an artificial gauge field :
a strong coupling perturbation approach}
\author{ Rashi Sachdeva, Sankalpa Ghosh }
\affiliation{Department of Physics, Indian Institute of Technology, Delhi, New Delhi-110016}
\begin{abstract}
We study the effect of an artificial gauge field  on the zero temperature phase diagram of
extended Bose Hubbard model, that describes ultra cold atoms in optical lattices with long range interaction
using strong coupling perturbation theory .  We determine analytically the  effect of the  artificial gauge field on the
density wave - supersolid (DW-SS) and the the Mott insulator-superfluid (MI -SF) transition boundary .
The momentum distribution at these two transition boundaries is also calculated in this  approach. It is shown that
such momentum distribution which can be observed in time of flight measurement, reveals the symmetry of the gauge potential through the formation of magnetic Brillouin zone and clearly distinguishes
between the DW-SS  and MI-SF boundary.  We also point out that in symmetric gauge the momentum distribution structure at these transition
boundaries bears distinctive signatures of vortices in supersolid and superfluid phases.
\end{abstract}
\pacs{03.75.Lm, 64.70.Tg, 67.80.bd}
\date{\today}
\maketitle

\section{Introduction}
The extended  Bose Hubbard Model (eBHM) incorporates the effect of long-range interaction upto various orders by adding
the interaction terms between bosons localized at different lattice sites and  has been widely studied \cite{Batrouni95,
otterlo, white, santos, pinaki, batrouni, pai, ramesh, lahaye}. Recently a number of  ultra cold atomic systems has been suggested as
possible candidates to realize such models. This includes condensate of ultracold dipolar  atoms like $^{52} Cr$ \cite{dipole1},
heteronuclear polar molecules \cite{polar1} and Rydberg excited ultra cold atomic condensate \cite{Rejish}.
The effect of long range interaction can be minimally taken care of by adding
a nearest neighbour interactions (NNI) in addition to onsite interactions to the prototype Bose Hubbard (BH) model \cite{BH}. The interest in this model stems from the
appearance of new phases, namely the Density Wave (DW) phase and the Supersolid (SS) phase, apart from the Mott Insulator (MI) and Superfluid (SF) phase. Both DW and MI phases lack phase coherence as the superfluid order parameter vanishes and, both these phases are incompressible with a finite gap in the particle-hole excitation spectrum.

For a system described by eBHM, in the limit where the hopping between neighbouring sites can be neglected in comparison to the onsite interaction, the system admits DW and MI phases alternatingly as the chemical potential increases. As the hopping amplitude
is increased in comparison to the onsite interaction and NNI the systems makes a transition from DW to a SS phase or from MI to a SF phase depending on the chemical potential. The  DW phase has alternating particle number at each site in contrast to the MI phase which has same number of particles at each site.
This characteristic of DW phase is interesting to explore as this checkerboard arrangement continues even when the interactions are reduced, resulting in transition to SS phase
prior to a SF phase.

In the SS phase \cite{Andreev, Legget, ODLRO} the SF and crystalline orders co-exist resulting in the spatial modulation of superfluid density.
This SS phase has been first claimed to be experimentally observed in solid Helium \cite{Mosechan}, however the interpretation of the experimental results is a matter of continuing debate  \cite{Reppy, review, Anderson, Balibar}. On the other hand, the cold atomic condensates with long range interactions loaded in an optical lattice can act as a relatively reliable way to confirm the existence of SS phase. Particularly the effect of artificial gauge field, that can be created either by means of rotation \cite{rotation, rotlat} or by imprinting motion-dependent laser induced phase on the internal states \cite{synthetic} on such systems, can lead to the formation of vortex states in such supersolid which has distinctive features as compared to the formation of similar vortices in an ordinary superfluid  both in terms of critical velocity of superflow \cite {Danshita} as well as profile \cite{rashi}.

One of the most common probing technique to detect the different phases in BH model or eBHM is to study their momentum distribution.
The information about the momentum distribution can be directly extracted from  time-of-flight (TOF) absorption imaging in the long time limit
of the freely expanding atoms released from the trap \cite{farfield, toth}. The experimental observation of MI-SF transition with ultra cold atoms in optical lattice \cite{Greiner} was based on this
idea. In this context in the current work we study the effect of artificial gauge field  on  insulating DW and MI phases and their respective transition to SS and SF phases using a strong coupling perturbation approach. We analytically calculate the modification of phase boundaries due to the gauge field, momentum distribution  and its gauge potential dependence. We also
 compare the momentum distribution of the vortex states carying definite quasi angular momentum at the the DW-SS and MI-SF boundary.

The effect of gauge field on the MI-SF transition in BH model has been studied extensively in recent times both within mean field  approximation
\cite{reijnder, rajiv, wu, goldbaum, Oktel, lundh, scarola}  and also by going beyond mean field description \cite{niemeyer, powell1, powell2, krishnendu}. Study of frustrated Bose-Hubbard model 
in presence of staggered fluxes using exact numerical diagonalization was also carried out recently and the dependence of the time of flight image on the gauge potential was discussed \cite{Moller}.
  In comparison, the DW - SS transition in eBHM in presence of a finite flux due to the gauge field \cite{rashi} as well as in presence of a staggered flux \cite{smith} is still in very early stage and was carried out  only in mean field framework.

The  strong coupling perturbation expansion which we adopt here to study the eBHM model  treats
the hopping as perturbation \cite{Freericks, Freericks1, Iskin} ( for an alternative way of doing strong coupling expansion  see \cite{dupuis})
and results from such approach matches quite well with results from quantum monte carlo simulation \cite{krauth}.
Particularly momentum distribution which can be compared with the experimental result of time of flight (TOF) imaging \cite{Krishnamurthy, Iskin2} can be calculated
 using strong coupling approach even
in the presence of an artificial gauge field where methods like quantum monte carlo is rather difficult to implement.

The remainder of the paper is organized as follows. In section II, we present the model Hamiltonian and the formalism of our calculations using the strong coupling perturbation theory. In section III, we present the analytical expressions for the phase boundaries between the incompressible (DW and MI) phases and compressible (SS and SF) phases, in presence of artificial gauge field. Section IV includes the calculation of momentum distribution in presence of artificial gauge field obtained by using wave function calculated by strong coupling perturbation theory expansion and also, the quasi angular momentum distribution of the states. A brief summary of the conclusions is presented in section V and the other details of the perturbation
calculations are provided in Appendix A and
B \\

\section{Extended Bose Hubbard Model in presence of magnetic field}
To describe the effect of nearest neighbour interaction on ultra cold atomic condensate loaded in a square optical lattice in the presence of
uniform artificial magnetic field in transverse direction, we introduce the following  extended Bose Hubbard Hamiltonian.
\beq H=-\sum_{i,j}t_{ij}\hat{b}_{i}^{\dag}\hat{b}_{j}+\frac{1}{2}\sum_{i}\hat{n}_{i}(\hat{n}_{i}-1)-\mu\sum_{i}\hat{n}_{i}
+V\sum_{i,j}\hat{n}_{i}\hat{n}_{j}\label{ham}\eeq
The first term gives us the nearest neighbour hopping where the hopping matrix elements are non zero only for nearest neighbours and is given by $t_{ij}=te^{i\phi_{ij}}$ with $\phi_{ij}= \int_{r_{j}}^{r_{i}}d\mathbf{r}.\mathbf{A}(\mathbf{r})$ and \ $\mathbf{A}(\mathbf{r})$ is the vector potential corresponding to the artificial gauge field. Here, $\hat{b}_{i}^{\dag}$, $\hat{b}_{i}$ and $\hat{n}_{i}$ are the boson creation, annhilation and number operators respectively. Here, V is the strength of nearest neighbour interaction that minimally captures the effect of long range interaction, $\mu$ is the chemical potential. We have rescaled the Hamiltonian by $U$ and thus, all parameters are measured in units of $U$. We neglect the effect of an overall trap potential assuming that it is sufficiently shallow and is neutralized by the effect of centrifugal force particularly at the central region of the condensate.
We are particularly interested in the limit when $Vd<U/2$, where $d$ is the dimension of the system which is $2$ in the
present case. In this limit, the alternating sites of the lattice  in the DW phase contains $n_{0}$ and $n_{0}-1$ particles and such a phase is called $n_{0}-\frac{1}{2}$ DW phase. Along the $t=0$ axis the system will form alternative sequence of
$n_{0}-\frac{1}{2}$ DW phases followed by MI phases of $n_0$ particles at each site. In the rest of the paper
that  the alternative sites of DW phase have population  $n_{A}$ and $n_{B}$ and finally set $n_{A}=n_{0}$ and $n_{B}=n_{0}-1$ to obtain the corresponding results for $n_{0} -\frac{1}{2}$  DW  phase.

\subsection{Formalism}
Within the frame work of the strong coupling perturbative expansion  we  calculate the ground-state energy $E_{DW}(n_{A},n_{B})$ and $E_{MI}(n_{0})$ of the DW phase with $n_{A}$ and $n_{B}$ bosons on alternating lattice sites, and of the MI phase with $n_{A}=n_{0}$ bosons on each lattice site, respectively. We also calculate the energies of the DW particle-hole excitations and MI particle-hole excitations (states with an extra particle or hole), $E_{DW}^{par}(n_{A},n_{B})$, $E_{DW}^{hol}(n_{A},n_{B})$, and $E_{MI}^{par}(n_{0})$, $E_{MI}^{hol}(n_{0})$, respectively.
The unperturbed system corresponds to the case $(t=0)$. In this limit, ground state energy as well as the particle hole
excitation energy can be analytically determined. Using the Rayleigh-Schroedinger perturbative expansion
the ground state energy of DW and MI phases as well the energy of particle hole like excitations over these ground state is calculated in various orders of  scaled  hopping parameter $t$.

Both the DW and MI states are gapped since the energy to create a single  particle-hole excitations is finite. With
increasing $t$ this gap energy starts decreasing. At the critical hopping parameter $t=t_c$  the energy to create a particle-hole pair vanishes and
DW phase becomes degenerate with its particle and hole excited state. This gives the value of $t$ at which DW-SS transition takes place. Thus, the phase boundary between the Density Wave and the Supersolid phase is determined by :
\beq E_{DW}(n_{A},n_{B})=E_{DW}^{par/hol}(n_{A},n_{B})\label{dw} \eeq
Similarly, the phase boundary between the Mott Insulator phase and the Superfluid phase is determined as :
\beq E_{MI}(n_{0})=E_{MI}^{par/hol}(n_{0})\label{mi}\eeq

These conditions determine the particle and hole branches of both the insulating lobes (DW and MI), giving us $\mu^{par}$ and $\mu^{hol}$ as functions of $ t, V, n_{A}, n_{B}$ (for DW phase) or $ t, V, n_{0}$ (for MI phase).\\

\subsection{Wave functions at zeroth order in $t$}
 In this section we shall define  the ground state wave functions for the DW and MI phases after setting  the scaled hopping amplitude  $t = 0$ for hamiltonian defined in (\ref{ham}).  In this limit these wavefunctions  are determined by the competition between interaction energies alone
and can be found out exactly. The wave functions for the particle and hole excited states  above these ground states will also be mentioned both for DW and MI phase and we shall particularly emphasize the degeneracy associated with such excited states in the $t \rightarrow 0$ limit.

For the DW state, we divide the lattice
into sublattices $A$ and $B$, where each site in sublattice $A$ contains $n_{A}$ particles and each site in sublattice $B$ contains $n_{B}$ particles. For $t=0$, DW wave function can be written as :
 \beq |\Psi_{DW}^{(0)}\rangle=\prod_{i\epsilon A, j\epsilon B}^{M/2}\frac{(\hat{b}_{i}^{\dag})^{n_{A}}}{\sqrt{n_{A}!}}\frac{(\hat{b}_{j}^{\dag})^{n_{B}}}{\sqrt{n_{B}!}}|0 \rangle\label{dwgd}\eeq
where $M$ is the total number of lattice sites, and $|0\rangle$ is the state with no particle. Here $\hat{b}_{i,j}^{\dag}$ refer to boson creation operator on $A$ and $B$ sublattices, respectively. Since we are interested in calculating the wave function as well as energy of such a state for finite $t$ as a perturbative
expansion in the parameter $t$, the wavefunction defined in (\ref{dwgd}) is also the wavefunction at the zeroeth order of this perturbative expansion.

Unlike the ground state wave function defined in (\ref{dwgd}), the wave functions for the DW states with an extra particle or hole for $t=0$ is degenerate. This is because
when an extra particle or hole is added to the system, it can go to any of the $M$ lattice sites. However in the case of a DW state, the alternatiing
sites belong to A and B sublattice and contain different number of particles. In the present case $n_{A}=n_{0}$ and $n_{B}=n_{0}-1$. Therefore a particle state over the DW ground state will consist of one particle added to any of site in sublattice
B, which contains less number of particle in the ground state. All such states as well their linear combination are degenerate for $t=0$. Since the total number of sites in B sublattice is $\frac{M}{2}$, therefore the dimension of this degenerate subspace of the one particle excitation is  also $\frac{M}{2}$. For
the single hole type of excitation over the DW ground state similarly the hole can be created in any site that belongs to sublattice A, containing higher
number of particle in the ground state. Therefore the dimension of the degenerate subspace of single hole like excitation is also $\frac{M}{2}$.
Because of this degeneracy, to find out the states and energies of particle and hole like excitation for finite $t$ as perturbative expansion in $t$ we need to use the degenerate perturbation theory.

To use degenerate perturbation theory  to find out the wave function as well as the energy for particle or hole like excited state, we need to diagonalize the perturbation hamiltonian.
  To do this we write $H$ given in (\ref{ham}) as
 \bea H = H_{0}+H_{P} \label{fullham}\eea
The unperturbed part,
\beq H_{0}=\frac{1}{2}\sum_{i}\hat{n}_{i}(\hat{n}_{i}-1)-\mu\sum_{i}\hat{n}_{i}+V\sum_{i,j}\hat{n}_{i}\hat{n}_{j}\label{half1ham} \eeq
and the perturbed part, which is the kinetic energy (hopping) term.
\beq H_{P}=-\sum_{i,j}t_{ij}\hat{b}_{i}^{\dag}\hat{b}_{j}\label{half2ham} \eeq
Now if we diagonalize $H_{P}$ in the degenerate subspace of either particle or hole like excitation over the DW ground state, we shall find that the degeneracy is only lifted when we include the second order ( next nearest neighbour) hopping
processes. This is because, each site in sublttice $A(B)$ has only the sites of sublattice $B (A)$ as its neighbour. Thus all the matrix element related by the first order hopping is $0$ and
we need to go upto the second order to lift the degeneracy. Here we briefly mention the methodology.

To find out the particle and hole excited state which will break the degeneracy when the perturbation is included we write it as a linear superposition of the degenerate basis states, namely
\bea |\Psi_{DW}^{par(0)}\rangle & = & \frac{1}{\sqrt{n_{B}+1}}\sum_{j\epsilon B}^{M/2}f_{j}^{DWB}\hat{b}_{j}^{\dag}|\Psi_{DW}^{(0)} \rangle\label{pardw} \\
 |\Psi_{DW}^{hol(0)}\rangle & = & \frac{1}{\sqrt{n_{A}}}\sum_{i\epsilon A}^{M/2}f_{i}^{DWA}\hat{b}_{i}|\Psi_{DW}^{(0)} \rangle\label{holdw}  \eea
The correct choice for $f_{j}$ and $f_{i}$ will be obtained by diagonalizing the second order perturbation due to the hopping matrix $t_{ij}$ and identifying the corresponding minimum eigenvalue.
Therefore  $f_{j}^{DWB}$ will be  the eigenvector of $\sum_{i}t_{ji}t_{ij^{'}}$ with the minimum eigenvalue ($\epsilon^{2}t^{2}$) such that $\sum_{i,j^{'}}t_{ji}t_{ij^{'}}f_{j^{'}}^{DWB}=\epsilon^{2}t^{2}f_{j}^{DWB}$ and $f_{i}^{DWA}$ will be  the eigenvector of $\sum_{j}t_{ij}t_{ji^{'}}$ with the minimum eigenvalue ($\epsilon^{2}t^{2}$) such that $\sum_{j,i^{'}}t_{ij}t_{ji^{'}}f_{i^{'}}^{DWA}=\epsilon^{2}t^{2}f_{i}^{DWA}$. The normalization condition also requires that $\sum_{j\epsilon B}^{M/2}|f_{j}^{DWB}|^{2}=1$ and $\sum_{i\epsilon A}^{M/2}|f_{i}^{DWA}|^{2}=1$.

Similarly for the  MI phase, the non degenerate ground state wave function for $t=0$ is given by:
\beq |\Psi_{MI}^{(0)}\rangle=\prod_{k=1}^{M}\frac{(\hat{b}_{k}^{\dag})^{n_{0}}}{\sqrt{n_{0}!}}|0 \rangle\label{migd}\eeq
where the index $k$ refers to all the lattice sites.
Here also,  for $t=0$ the single particle or hole like excited states over this non-degenerate ground state is degenerate and dimension of these degenerate subspace is $M$, the total number of lattice sites. Since all sites are equivalent, this degeneracy is lifted in the first order correction of the degenerate perturbation theory,  namely when the nearest neighbour hopping is included, unlike the previous case in DW phase, where to lift the degeneracy one needs to include the next nearest neighbour hopping. Here very briefly we define the details.

The single particle and hole excited state wave functions that will break the degeneracy for finite $t$ is written as
\bea |\Psi_{MI}^{par(0)} \rangle & = & \frac{1}{\sqrt{n_{0}+1}}\sum_{k=1}^{M}f_{k}^{MI}\hat{b}_{k}^{\dag}|\Psi_{MI}^{(0)} \rangle \label{miparwf} \\
 |\Psi_{MI}^{hol(0)} \rangle & = & \frac{1}{\sqrt{n_{0}}}\sum_{k=1}^{M}f_{k}^{MI}\hat{b}_{k}|\Psi_{MI}^{(0)} \rangle \label{miholwf} \eea
The correct choice of the $f_k$ will be obtained by diagonalizing the first order perturbation due to hopping  and locating the minimum eigenvalue state.
Thus $f_{k}^{MI}$ is the eigenvector of the hopping matrix $t_{kk^{'}}$ with the minimum eigenvalue ($\epsilon t$)

It may be noted that the $H_{P}$ is same as the Harper hamiltonian whose spectrum is given by the Hofstadter butterfly. Therefore
finding the  solution that corresponds to the minimal eigenvalue  of the hopping matrix $t_{ij}$ is identical to finding the band minimum in the Hofstadter problem
and to locate their corresponding eigenstates.  In the following section we shall provide the analytical expression of the perturbatively calculated energy of the ground state
and the particle-hole excitation at finite $t$ and evaluate the phase boundary of the DW-SS transition and the MI-SF transition from this result.

\section{Analytic expressions for the modification of phase boundaries}
Using  many body version of the  Rayleigh-Schroedinger perturbation theory
 the ground state energy of the insulating phases ( DW or MI)  as well as the energies of the particle or hole like excitation can be expressed as  a power series in  the scaled hopping amplitude $t$

 \beq  E (t)=t^{0}E_{n}^{(0)}+t^{1}E_{n}^{(1)}+t^{2}E_{n}^{(2)}+t^{3}E_{n}^{(3)} + \cdots \label{enins}\eeq
 where $E_{n}^{(0)}$ is the energy in the limit $t=0$, and $tE_{n}^{(1)}$, $t^2 E_{n}^{(2)}$ and $t^3 E_{n}^{(3)}$ are the first order, second order and third order corrections to energy. In the following we present such calculation upto the third order in $t$  and the other details associated with the calculation are relegated to the Appendix A

The expression of the ground state energy of the DW wave state per site is given as
\begin{widetext}
\bea \frac{E_{DW}^{ins}}{M}=\left [\frac{1}{2}(n_{0}-1)^{2}+zV\frac{n_{0}(n_{0}-1)}{2}-\mu\frac{2n_{0}-1}{2} \right]
-z\left[\frac{n_{0}^{2}}{V(z+1)}+\frac{n_{0}^{2}-1}{2-V(z+1)}\right]t^{2}+O(t^{4})\label{endwgd}\eea
\end{widetext}
In the above expression $z=2d$ is the co-ordination number of given lattice site which is in the case of a square lattice $4$. The first term is the ground state energy  at $t=0$ and
the second term is the second order correction due to finite $t$. Both the first and third order correction vanishes,
We now calculate the energies of the  states with an extra particle or hole using Eqs.(\ref{pardw}) and (\ref{holdw})  in the framework of degenerate perturbation theory.
For the particle excited DW states,

\begin{widetext}
\bea E_{DW}^{p} &=& {E_{DW}^{ins}}+ t^{0}[(n_{0}-1)+zVn_{0}-\mu] \nonumber \\
                          &  &  \mbox{} + t^{2} [ -\frac{n_{0}(n_{0}+1)\epsilon^{2}}{(1-zV)}+
\frac{n_{0}^{2}\epsilon^{2}}{((z-1)V)}
 +\frac{(n_{0}^{2}-1)\epsilon^{2}}{(2-(z+1)V)}
 -\frac{n_{0}(n_{0}+1)z}{(1+(z-2)V)}\nonumber\\
  &  & \mbox{} -\frac{(n_{0}^{2}-1)(\epsilon^{2}-z)}{(2-zV)}+
\frac{2n_{0}^{2}(\epsilon^{2}-z)}{((z-2)V)} ]  + O(t^{4}) \label{endwpar} \eea
\end{widetext}
Comparing this expression with Eq. \ref{enins} we find that the first and third order correction vanishes.
 In a similar manner the energy of the hole excited DW states can be found out. This turns out to be
\begin{widetext}
\bea  E_{DW}^{h} &=& {E_{DW}^{ins}}+ t^{0} [-(1+zV)(n_{0}-1)+\mu ] \nonumber \\
                              &  & \mbox{} + t^{2}[-\frac{n_{0}(n_{0}-1)\epsilon^{2}}{(1-zV)}+
\frac{n_{0}^{2}\epsilon^{2}}{((z-1)V)}
 -\frac{(n_{0}^{2}-1)\epsilon^{2}}{(2-(z+1)V)} \nonumber\\
 &  & \mbox{}+\frac{n_{0}(n_{0}-1)z}{V(1+(z-2)V)}
-\frac{(n_{0}^{2}-1)(\epsilon^{2}-z)}{V(2-zV)}+
\frac{2n_{0}^{2}(\epsilon^{2}-z)}{(1+(z-2)V)} ]
+O(t^{4}) \label{endwhol}\eea
\end{widetext}

In a similar way, the expression for energy of the MI phase and a single particle or hole excitation over it as  perturbative expansion in the scaled hopping parameter $t$ upto third order is given by
\begin{widetext}

\beq \frac{E_{MI}^{ins}}{M} = \frac{1}{2}n_{0}(n_{0}-1)-\mu n_{0}+zV\frac{n_{0}^{2}}{2}
  -zt^{2}\frac{n_{0}(n_{0}+1)}{(1-V)}+O(t^{4})\label{enmigd}\eeq

\bea  E_{MI}^{p} & =  & {E_{MI}^{ins}}+t^{0} [n_{0}+zVn_{0}-\mu] \nonumber \\
&  & -t^{1} (n_{0}+1)\epsilon +  t^2[n_{0}(n_{0}+1)\left[(z-\epsilon^{2})+\frac{2(z-\epsilon^{2})}{(1-2V)}+\frac{2\epsilon^{2}}{(1-V)}\right]
-n_{0}(n_{0}+2)\frac{z}{2(1-V)}] \nonumber \\
& & \mbox{} + t^{3}[-n_{0}(n_{0}+1)n_{0}\left[(z-2\epsilon)\epsilon+\frac{(\epsilon^{2}-3z+3)\epsilon}{(1-V)^{2}}\right]
\nonumber\\
&  & \mbox{}- n_{0}(n_{0}+1)(n_{0}+1)\left[(z-\epsilon^{2})\epsilon-\frac{(2\epsilon^2-6z+6)\epsilon}{(1-V)^{2}}+
\frac{2\epsilon(z-\epsilon^{2})}{(1-2V)^{2}}+\frac{2\epsilon(\epsilon^{2}-3z+3)}{(1-V)}+\frac{4(z-2)\epsilon}{(1-2V)}
+\frac{4\epsilon(\epsilon^{2}-3z+3)}{(1-V)(1-2V)}\right]\nonumber\\
&  & \mbox{} -n_{0}(n_{0}+1)(n_{0}+1)\frac{4\epsilon(\epsilon^{2}-3z+3)}{(U-V)(U-2V)}
-n_{0}(n_{0}+1)(n_{0}+2)
 \left[\frac{\epsilon(z-1)}{(1-V)}
-\frac{z\epsilon}{4(1-V)^{2}}\right]]  + O(t^{4})\label{enmipar}\eea

\bea E_{MI}^{h}  & = & E_{MI}^{ins}+ t^{0}(-(n_{0}-1)-zVn_{0}+\mu)  \nonumber \\
& & \mbox{} -t[n_{0}\epsilon ] + t^2[ (n_{0}+1)n_{0}\left[(z-\epsilon^{2})+\frac{2(z-\epsilon^{2})}{(1-2V)}+\frac{2\epsilon^{2}}{(1-V)}\right]
-(n_{0}+1)(n_{0}-1)\frac{z}{2(1-V)}] \nonumber \\
&   & \mbox{} t^{3}[-n_{0}(n_{0}+1)(n_{0}+1)\left[(z-2\epsilon)\epsilon+\frac{(\epsilon^{2}-3z+3)\epsilon}{(1-V)^{2}}\right]
\nonumber\\
&  & \mbox{} - n_{0}(n_{0}+1)n_{0}\left[(z-\epsilon^{2})\epsilon-\frac{(2\epsilon^2-6z+6)\epsilon}{(1-V)^{2}}+
\frac{2\epsilon(z-\epsilon^{2})}{(1-2V)^{2}}+\frac{2\epsilon(\epsilon^{2}-3z+3)}{(1-V)}+\frac{4(z-2)\epsilon}{(1-2V)}
+\frac{4\epsilon(\epsilon^{2}-3z+3)}{(1-V)(1-2V)}\right]\nonumber\\
&  & \mbox{} -n_{0}(n_{0}+1)(n_{0}+1)\frac{4\epsilon(\epsilon^{2}-3z+3)}{(U-V)(U-2V)}-
-n_{0}(n_{0}+1)(n_{0}-1)
 \left[\frac{\epsilon(z-1)}{(1-V)}
-\frac{z\epsilon}{4(1-V)^{2}}\right]]   + O(t^{4})  \label{enmihol} \eea
\end{widetext}

It may be seen that unlike in the case of DW phase, for the MI phase the the perturbative correction to the energy for finite $t$ appears in the first order term itself and
the second and third order corrections that appear in Eq.( \ref{enmipar}) and  (\ref{enmihol}) is comparatively much smaller in magnitude. However we still kept the calculation upto
the same order so that the entire phase diagram which comprises of DW as well as MI phase is given within the same order of the perturbation theory.
In the following sections, using  from the above expressions of energy of the ground state and their corresponding particle hole excitation we shall determine the MI-SF and DW-SS phase
boundary from the relations  (\ref{dw}) and (\ref{mi}).

\subsection{Determination of the DW-SS and MI-SF boundary from the strong coupling expansion}
\begin{figure}[ht]
\centerline{ \epsfxsize 10cm \epsfysize 8cm \epsffile{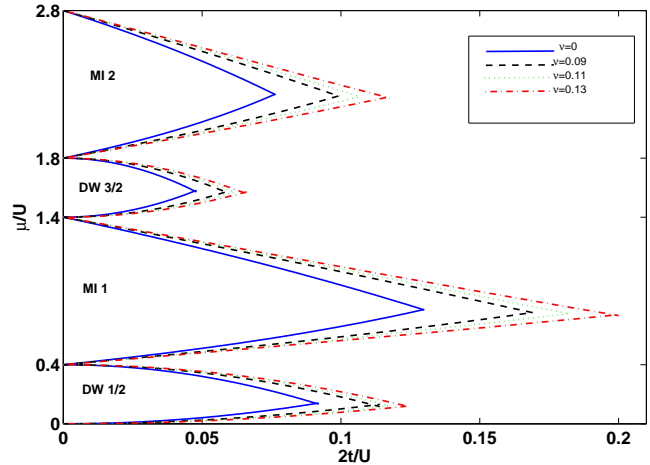}}
 \caption{\textit{(Color Online)} Phase diagram for the eBHM in presence of artificial magnetic field for Vd/U=0.2. The effect of increasing magnetic field is to increase the stability of the DW and MI lobes }
 \label{ebhmrot}
\end{figure}

Fig. (\ref{ebhmrot}) represents the phase diagram of the first four insulating lobes for the eBHM in two dimension (square lattice) in presence of artificial magnetic field. It shows the increasing stability of the insulating phase i.e the DW and MI phases grows in size as the strength of magnetic field is increased from zero to finite values. This is due to the localizing effect of magnetic field on the moving bosons, thus favouring the insulating phases to occupy a larger area in the phase diagram.

The minimum eigen value $\epsilon$ used in calculation of energy expressions, involves the diagonalization of the hopping matrix, which is gauge dependent. The location of the minimum eigen value $\epsilon$ depends on the choice of the gauge potential, while the eigen value itself is not. We consider a rational flux $\nu=p/q$ and calculate the minimum eigen values of the hopping matrix $t_{ij}$ following \cite{kohmoto}.
The shape of the insulating lobes for the Mott phase is different from the DW phase, as observed in Fig (\ref{ebhmrot}) in two dimensions. This is because, the Mott states with an extra particle or hole, have corrections to first order in $t$ , while the DW states have corrections to second order in $t$. So, as $t\rightarrow 0$, the slope of the Mott state will be finite, while it will vanish for the DW lobes. Hence, the shapes of the two insulating lobes are always different. Moreover, the shapes of the insulating lobes depend on the dimensionality of the system and also, on the application of artificial magnetic field \cite{niemeyer}. The mean field theories always give a concave shape to the MI and DW lobes\cite{rashi} as the dimensionality only enters as a prefactor in the expression for $\mu$ as a function of $t$, while the strong coupling expansion easily distinguishes the shape of insulating lobes in different dimensions and also, in presence of artificial magnetic field.
The determination of these transition boundary using strong coupling perturbation theory is one of the main results in this paper,
Since the transition at the boundary in this model belongs to the universality class of $2+1$ dimensional XY model, the critical exponents can also be found out through an expansion of
the chemical potential. This was done in \cite{Iskin} and we shall not discuss this issue further.

In the next section, we calculate the effect of artificial magnetic field on the momentum distribution of the insulating phases (DW and MI), with both gauge choices, Landau gauge and symmetric gauge.

\section{Calculation of Momentum distribution}
The standard experimental way of probing  the properties of an ultra cold atomic condensate is through TOF absorption imaging of the freely expanding atoms released from the trap \cite{toth}. Same method is used for probing the condensate loaded
in optical lattice as well. The quantity that is measured experimentally in such TOF expansion is the momentum distribution
of this ultra cold atomic ensembles.
In this section we shall first provide the method of calculation of the momentum distribution within the framework of strong coupling expansion. Subsequently we provide our results and their analysis. We shall particulary show that in the presence
of an artificial magnetic field the TOF immage actually depends on the means to produce effective magnetic field.
As already mentioned in all these discussion we have not taken into account the existence of an overall trap potential assuming it to shallow. Thus the discussion  we present can be related to the bulk region of  trapped condensate where
the effect of the trap potential can be neglected.

The momentum distribution $n (\mathbf{k})$  is defined as the Fourier transform of the one body density matrix, $\rho(\mathbf{r}, \mathbf{r^{'}})=\langle \psi^{\dag}(\mathbf{r})\psi(\mathbf{r^{'}}) \rangle$, \cite{pitstri}. Therefore,
\beq n(\mathbf{k})=\int d\mathbf{r}\int d\mathbf{r^{'}}\rho(\mathbf{r},\mathbf{r^{'}})e^{i\mathbf{k}.(\mathbf{r}-\mathbf{r^{'}})}  \label{mdf} \eeq
where $ \psi^{\dag}(\mathbf{r})$ and $\psi(\mathbf{r})$ are the bosonic field operators at the position $\mathbf{r}$, respectively, and $\mathbf{k}$ is the momentum. The expectation value is taken in the many-body ground state and the corresponding wave function is determined using strong coupling perturbation theory as a power series in  scaled hopping amplitude $t$ ( in the unit of $U$).

To evaluate (\ref{mdf}) we expand the field operators in the basis set of Wannier functions, such that
\beq \psi(\mathbf{r})=\frac{1}{\sqrt{M}}\sum_{l}w(\mathbf{r}-\mathbf{R}_{l})b_{l}
\nonumber\eeq 
where $M$ is the total number of lattice sites, $w(\mathbf{r}-\mathbf{R}_{l})$ is the Wannier function localized at site $l$ with position $R_{l}$ and $b_{l}$ is the boson annihilation operator. Consequently, the momentum distribution becomes
\beq n(\mathbf{k})=\frac{1}{M}\sum_{l,l^{'}}w^{*}(\mathbf{k}_{l})w(\mathbf{k}_{l})\langle b_{l}^{\dag}b_{l^{'}} \rangle e^{i\mathbf{k}.
(\mathbf{R}_{l}-\mathbf{R_{l^{'}}})}\label{nk}\eeq
where $w(\mathbf{k})=\int d\mathbf{r}w(\mathbf{r})e^{i\mathbf{k}.\mathbf{r}}$ is the fourier transform of the usual Wannier function $w(\mathbf{r})$. The summation indices $l\epsilon\left\{A,B\right\}$ and $l^{'}\epsilon\left\{A,B\right\}$ includes the entire lattice.   The wave functions for the insulating phases are calculated perturbatively upto second orer in $t$ which is the first significant order.
The higher order correction can be neglected since $t \ll 1$,
Upto second order, the wave function for the insulating state (MI/DW) can then be written as
\beq |\psi_{ins} \rangle =|\Psi_{ins}^{(0)} \rangle +|\Psi_{ins}^{(1)} \rangle +|\Psi_{ins}^{(2)} \rangle+O(t^{3})
\label{genwf} \eeq
Here,  $| \Psi_{ins}^{0} \rangle=|\Psi_{MI/DW}^{(0)} \rangle$  defined in (\ref{dwgd}) and (\ref{migd})
and \bea |\Psi_{ins}^{(1)} \rangle & = & \sum_{m\neq|\Psi_{ins}^{(0)} \rangle}\frac{\langle m|H_{P}|\Psi_{ins}^{(0)} \rangle}{E_{0m}}|m \rangle \label{1stint} \\
 |\Psi_{ins}^{(2)} \rangle & = & \sum_{m,m^{'} \neq|\Psi_{ins}^{(0)} \rangle}\frac{ \langle m^{'}|H_{P}|m\rangle \langle m|H_{P}|\Psi_{ins}^{(0)} \rangle }{E_{0m^{'}}E_{0m}}|m^{'} \rangle \nonumber \\
&   &  \label
{2ndint} \eea
In the expression  for the first order correction (\ref{1stint}), the matrix element of  $H_{P}$ (hopping matrix)
is taken between the first-order intermediate (excited) state $|m>$ and the zeroth-order state defined in (\ref{dwgd}) and
(\ref{migd}), where as $|m^{'}\rangle$ in the expression of the second order correction (\ref{2ndint}) is the second order intermediate (excited) state.  Also, $E_{0m}= E_{0}-E_{m}$ is the energy difference between the first order intermediate state $|m\rangle$ and zeroth order state $|\Psi_{ins}^{(0)}>$ and $E_{0m^{'}}= E_{0}-E_{m^{'}}$ is the energy difference between the second order intermediate state $|m^{'}>$ and zeroth order state $|\Psi_{ins}^{(0)} \rangle$  \\

\begin{figure}[ht]
\centerline{\epsfxsize 8cm \epsfysize 6.5cm \epsffile{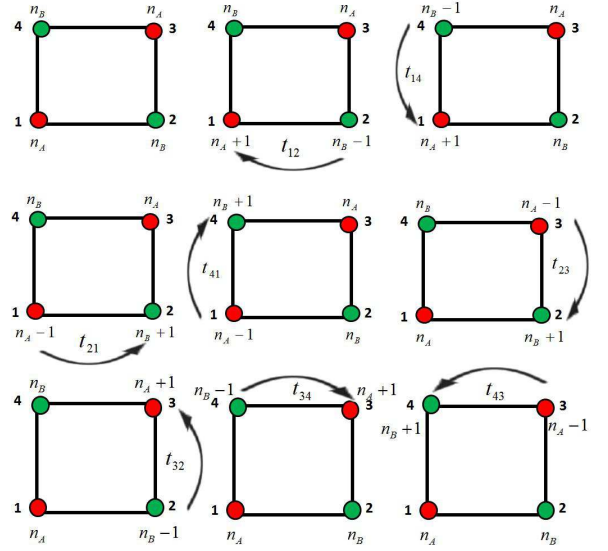}}
\caption{\textit{(Color Online)} Possibilities for the first order intermediate (excited) state for the DW phase in a 2x2 lattice. The figure (a) corresponds to the
ground state wave function in the limit $t=0$}
\label{1stexcited}
\end{figure}

\begin{figure}[ht]
\centerline{ \epsfxsize 8cm \epsfysize 8cm \epsffile{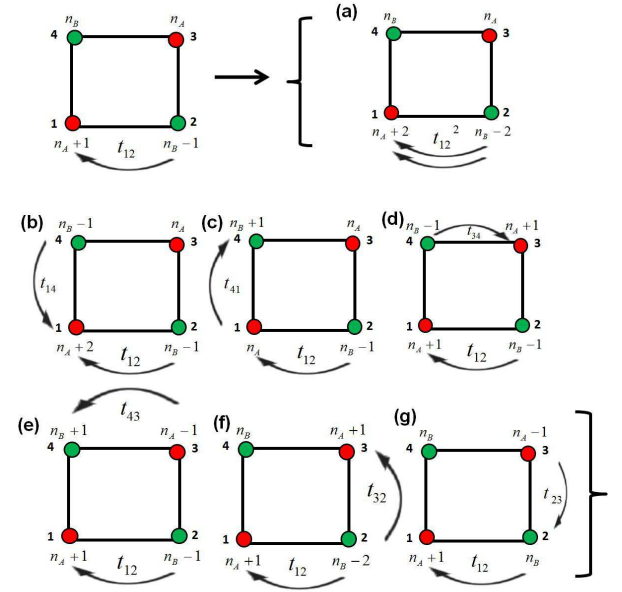}}
\caption{\textit{(Color Online)} Second order intermediate states for $|m>=|n_{A}^{(1)}+1, n_{B}^{(2)}-1, n_{A}^{(3)}, n_{B}^{(4)} \rangle$ ( the first diagram)}
\label{2ndexcited}
\end{figure}

To elucidate the meaning of the above terms appearing at the various orders of the perturbation theory, we depict the first and second order hopping processes respectively in Fig. (\ref{1stexcited}) and Fig. (\ref{2ndexcited})
for the simplest case of a $2 \times 2$ square lattice unit cell.
In the diagram given in Fig. (\ref{1stexcited}) (a),  we depict the ground state wave function for the DW state for $t=0$
for this unit cell whose wave function is
can be written as
\beq |\Psi_{DW}^{(0)}\rangle= |n_{A}^{(1)},n_{B}^{(2)},n_{A}^{(3)},n_{B}^{(4)} \rangle \label{dwwfgd} \eeq
 Here the superscript $(i)$ correspond to the location of a given lattice site.
All the subsequent diagrams in the same figure (Fig.\ref{1stexcited}$(b)$-$(i)$) depict all the possible $| m \rangle$ state defined in (\ref{1stint}) which is connected
to the DW ground state in Fig. \ref{1stexcited}(a)  by a single hopping.  For example,
the state depicted in  Fig. (\ref{1stexcited})  correspond to $|m \rangle = |n_{A}^{(1)}+1, n_B^{(2)} -1, n_{(A)}^{3}, n_{(B)}^{4} \rangle$ .
The matrix element of $H_P$ between this state and the ground state can be calculated as
\beq \frac{\langle m|H_{P}|\Psi_{ins}^{(0)} \rangle
}{E_{0m}}=\frac{-\sqrt{n_{B}(n_{A}+1)}t_{12}}{(-2+(z+1)V)}\nonumber\eeq
and contribute to the first order corrections in the expression (\ref{1stint}).
The further details of  these calculations
are given in the Appendix B.

Next, we calculate the second order correction to the wave function of the DW state (eq \ref{2ndint}). For every first order intermediate state $|m\rangle$, there will be number of possible second order intermediate states.
For example, for the just discussed  $|m\rangle=|n_{A}^{(1)}+1, n_B^{(2)} -1, n_{A}^{(3)}, n_{B}^{(4)} \rangle$ (Fig.\ref{1stexcited}$(b)$)
which is connected to the ground state by single hooping,
the corresponding possibilities for $|m^{'}\rangle$  are shown in Fig. (\ref{2ndexcited} (a)-(g)) which is
seven in number.

In the same way, for each possible first order intermediate (excited) state $|m\rangle$ (Fig. \ref{1stexcited}$(b)$-$(i)$), the second order excited states are calculated and thus by adding all possible combinations, the second order correction to wave function is determined (Eq. \ref{2ndint}). The details of the calculation is provided in Appendix B.

With the help of above expressions for perturbation corrections, the DW ground state wave function can be
determined  upto the second order in perturbation theory using the general expression (\ref{genwf})
and subsequently normalized within the same order of perturbation theory. Substituting of this wavefunction in the
expression (\ref{nk}) for  momentum distribution yields th $n(\bs{k})$ for the DW phase as
\begin{widetext}
\bea n_{DW}(\mathbf{k}) & =  & \frac{2n_{0}-1}{2}
-\left[\frac{n_{0}^{2}}{(z-1)V}
+\frac{(n_{0}^{2}-1)}{2-(z+1)V}\right]\epsilon(\mathbf{k})\nonumber\\
&  & \mbox{} + 2n_{0}\left[\frac{n_{0}^{2}}{2(z-1)^{2}}
V^{2}+\frac{(n_{0}^{2}-1)}{2[2-(z+1)V]^{2}}
+\frac{n_{0}^{2}}{(z-1)V}+\frac{(n_{0}^{2}-1)}{[2-(z+1)V]}\right][\epsilon^{2}(\mathbf{k})-2dt^{2}] \nonumber \\
&  & \mbox{} +O(t^{3})
\label{momdw} \eea
\end{widetext}

Similarly, for the MI phase with $n_{0}$ particles on each lattice site, the momentum distribution in presence of gauge field is
\bea n_{Mott}(\mathbf{k})& = & n_{0}-\frac{2n_{0}(n_{0}+1)}{1-V}\epsilon(\mathbf{k})+n_{0}(n_{0}+1)(2n_{0}+1)\nonumber\\
&   & \times(\epsilon^{2}(\mathbf{k})-2dt^{2})\frac{3-2V}{(1-V)^{2}}+O(t^{3})\label{mommi}\eea

In either of the above expression of
 momentum distribution (\ref{momdw})  and (\ref{mommi}), the dispersion
$\epsilon (\bs{k})$ is the minimum eigenvalue of the artificial magnetic flux dependent hopping matrix $T$ or $t_{ij}$
 multiplied by a pre-factor $\frac{2}{M}$ for DW phase and $\frac{1}{M}$ for the MI phase where $M$
is the total number of sites along a given direction. The matrix $T$
in the lattice site basis can be written
 \beq T=
 \left[ \begin{array}{ccccc}
t_{11} & t_{12} & t_{13} & \ldots & t_{1M}\\
t_{21} & \ldots & t_{ij} &\ldots&  \\
\vdots &\vdots & \vdots & \vdots & \vdots\\
t_{m1} & \ldots & & t_{M,M-1}& t_{MM} \\
 \end{array} \right ] \label{tmatreal}\eeq
As described in  (\ref{ham}), $t_{ij}=te^{i \phi_{ij}}$ is the gauge  dependent
hopping amplitude from site $i$ to site $j$  and non-vanishing only  if $i$ and $j$ are nearest neighbours. Through  $ \phi_{ij}=\int_{r_{j}}^{r_{i}}d\mathbf{r}.A(\mathbf{r})$  this hopping amplitude
explicitly depend on the gauge
potential with $A(\mathbf{r})$ as the vector potential.
In the next sub section, we provide the results for the effect of artificial magnetic field on the momentum distribution of the DW and MI phases.
\subsection{Effect of the presence of gauge field on the momentum distribution}\label{epsexp}
The effect of artificial gauge field on the
on the  MI-SF and DW-SS transition boundary have distinctive features which can be demonstrated
by plotting the momentum distribution derived in (\ref{momdw}) and (\ref{mommi}) in the $k_x-k_y$ plane at these transition boundaries.  Since the matrix $T$ in (\ref{tmatreal}) explicitly depends on the gauge potential,
the momentum distribution
reflects in itself the gauge potential structure.

To demonstrate this  we calculate the dispersion $\epsilon(\bs{k})$ for  two types of artificial gauge fields on the system, the Landau gauge and the Symmetric gauge and evaluate the corresponding momentum distribution structure.
The vector potential in Landau ($^{L}$) and Symmetric ($^S$)   gauge are respectively given as
\bea  A^L(\mathbf{r}) & = & \mathbf{B}(0,x,0) = 2 \pi \nu x \hat{y}  \label{landau1} \\
A^{S}(\mathbf{r}) & = & \mathbf{B}(-y,x,0)=\pi\nu(x\hat{y}-y\hat{x})  \label{symm1}\eea
 where we consider that  the flux due to the corresponding gauge field through the unit shell  $\nu=p/q$  is rational.

In the Landau gauge $\mathbf{A}^{L}$, following \cite{kohmoto}
we denote the co-ordinate of a site $i$ on the square lattice by  pair of integers  $\{m,n \}$  in  the unit of the
lattice spacing $a$. For Landau gauge potential therefore the phase of the hopping parameter $\phi_{ij}=0$ is if the
link along $x$ direction ( $ i \rightarrow j = \{ m,n \} \rightarrow  \{ m+1, n \}$) and $\phi_{ij}=2 \pi  n \nu $ if
the link  is along the $y$ direction such ( $ i \rightarrow j = \{ m,n \} \rightarrow  \{ m, n+1 \}$ ).  In terms of these notations the eigenvalue equation $T \psi= \epsilon \psi$
on the lattice can be written as
\bea \lefteqn{ -t [\psi _{n+1,m} + \psi_{n-1, m}  + {} } \nonumber \\
&  &  e^{ 2 \pi i \nu n} \psi_{n, m+1} + e^{-2 \pi i \nu n} \psi_{n, m-1}]=\epsilon
\psi_{n, m} \label{harper1} \eea

The lattice wavefunctions that appears in  Eq. (\ref{harper1})
can be obtained by operating $\psi_{n,m}$ by  magnetic
translation operator. Such operators are given as  $T_{\bs{R}}=exp (  \frac{i}{\hbar} \bs{R} \cdot [\bs{p} + \frac{h}{m} A(\bs{r})])$ \cite{Zak} where $R$ is the lattice translational vector and it is known
the operators along the $x$ and $y$ axis do not commute since $ \hat{T}_{a \hat{x}} \hat {T}_{a \hat{y}}
\hat{T}_{a \hat{x}}^{-1} \hat{T}_{a \hat{y}}^{-1}
= \exp ( 2 \pi i \nu)$. In case of Landau gauge and for $\nu = \frac{p}{q}$  the required
commutator is given by $ [T_{qa \hat{x}}, T_{a \hat{y}}]=0$. To ensure that the wave function remain single valued
at a given lattice point as an unit cell is traversed, the enlarged unit cell known as magnetic unit cell therefore has  $q
\times 1$ sites.
as compared to the unit cell in the absence of such magnetic field.
 Correspondingly  the Brillouin Zone (BZ) is reduced with $-\pi\leq k_{y}\leq\pi, -\pi/q \leq k_{x}\leq \pi/q$  and is
called the magnetic Brillouin zone (MBZ). As we shall see the momentum distribution $n (\bs{k})$ shows the formation
of such MBZ.

\begin{figure}[ht]
\centerline{ \epsfxsize 7cm \epsfysize 6cm \epsffile{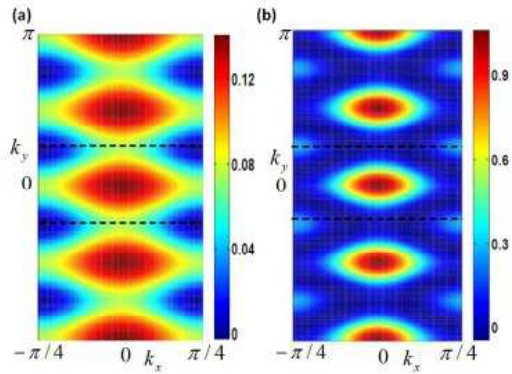}}
\caption{\textit{(Color Online)} Momentum distribution for the (a) MI and (b) DW phase in landau gauge potential for $\nu=\frac{1}{4}$, plotted in the reduced brillouin zone $-\pi\leq k_{x}\leq \pi$
and $-\pi/4\leq k_{y}\leq \pi/4$}
\label{landaucomp}
\end{figure}

Since $ [T , \hat{p}_{y}]=0$ in the Landau gauge the wave function $\psi_{n,m}( k_x , k_y) = e^{i k_y m} e^{ i k_x n}
\phi_{n} (k_x, k_y)$ where $k_{x,y}$ are the components of the Bloch wave vectors. Substituting this expression in the eigenvalue
equation (\ref{harper1}) one gets the following one dimensional eigenvalue equation
 \beq -t[ e^{i k_x} \phi_{n+1} + e^{-i k_x } \phi_{n_{1}} + 2 \cos ( k_y + 2 \pi \nu n) \phi_{n} ] = \epsilon \phi_{n} \eeq
The eigenvalues $ \epsilon ( \bs{k})$ now can be determined  from the condition

\beq  \det \left[ \begin{array}{ccccc}
M_{1} & e^{-i k_x }& 0 & \ldots & e^{i k_x}\\
e^{ i k_x} & M_{2} & \ldots &\ldots&  \\
0 &\vdots & \vdots & \vdots & 0\\
\ldots & \ldots & & M_{q-1}&e^{-i k_x}\\
N_{q} & & 0 & e^{ i k_x}& M_{q} \\
 \end{array} \right ]  = 0 \eeq

with $M_{n}= 2cos(k_{y}a+2\pi n \phi)-\epsilon(\bs{k})$ and, where $n=0,1,...,q-1$.
The eigenvalue matrix  is $ q \times q$ dimensional because of the
q times enhancement of the magnetic unit cell. And its minimum
eigenvalue $\epsilon( \bs{k})$ is $q$ -fold degenerate since $ \epsilon ( k_x , k_y) = \epsilon(k_x , k_y + \frac{ 2 \pi \alpha}{q})$ where $ \alpha = 0,  \cdots , (q-1)$ . Each of this minimum will correspond to a peak in the momentum distribution
which is calculated using (\ref{momdw}) and (\ref{mommi}). This has been plotted in  Fig (\ref{landaucomp}(a)) and (b)
for MI and DW phase.

\begin{figure}[ht]
\centerline{ \epsfxsize 10cm \epsfysize 5cm \epsffile{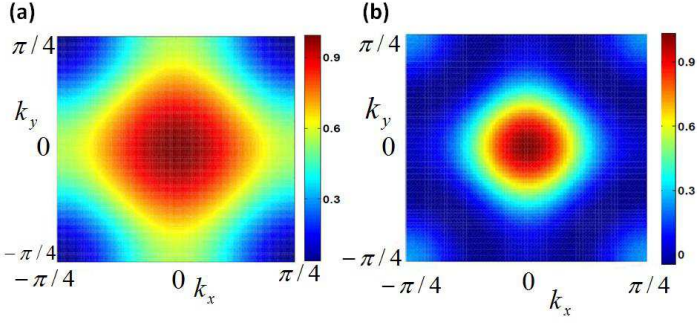}}
\caption{\textit{(Color Online)} Momentum distribution for the (a) MI and (b) DW phase in symmetric gauge potential for $\nu$=1/4, plotted in the range $-\pi/4\leq k_{x}\leq \pi/4$
and $-\pi/4\leq k_{y}\leq \pi/4$}
\label{symmcomp}
\end{figure}

 Thus in the Landau gauge potential, in MI phase one gets $q$ peaks in $n(\mathbf{k})$ at $(k_{x},k_{y})=(0, \frac{2\pi n}{q})$, with $n=0,1,..q-1$. Accordingly
Fig (\ref{landaucomp}(a)) shows the momentum distribution for $q=4$  which has  peaks at $(0,0)$, $(0, \pm\pi/2)$ and $(0, \pm\pi)$.
For the DW phase, $\nu=1/q$ with $q=4$, we again get peaks at $n(\mathbf{k})$ at $(0,0)$, $(0, \pm\pi/2)$ and $(0, \pm\pi)$, but in addition, there
also exists small peaks at the corners of the reduced BZ, as shown in fig (\ref{landaucomp}(b)). The appearance of extra peaks at the BZ corners  distinguishes  DW phase from the MI phase. Since the DW phase consists of two interpenetrating sublattices $A$ and $B$,  even in the absence of any applied magnetic field, the BZ structure of the DW phase has double
 periodicity whereas the usual MI state  has a  single periodicity. This results in appearance of small extra peaks in the momentum distribution of DW even in the absence of magnetic field \cite{Iskin2}. Upon application of magnetic field, we observe small peaks in the DW momentum distribution at the BZ corners, which is again attributed to the reduced periodicity of the DW phase compared to MI phase.\\
When one uses a symmetric gauge potential  given in (\ref{symm1}) the commutating magnetic translation operators are $\hat{T}_{2qa \hat{x}}$ and $\hat{T}_{2qa \hat{y}}$. The corresponding discretized eigenvalue equation was discussed in detail in
\cite{rashi}. Following the preceeding discussion it can be shown that the magnetic unit shell should be $ 2q \times 2q$
of the unit cell in the absence field. The factor $2$ comes because the phase accumulated when one goes around
an unit cell in the square lattice is $\pi i \nu$  in presence of the gauge potential (\ref{symm1}).
Accordingly the MBZ will be defined as  $(-\frac{\pi}{2q} \leq k_{x}\leq \frac{\pi}{2q}, -\frac{\pi}{2q}\leq k_{y}\leq \frac{\pi}{2q})$. This results  in the formation of peaks in momentum distribution at $(\pm \pi n/q,\pm \pi n/q)$.
The momentum distribution of the MI phase in presence of symmetric gauge potential, for rational flux $\nu=1/4$ is shown in fig (\ref{symmcomp}(a)). The peak in the momentum distribution is observed at  the center ($(0,0)$) and smaller peaks are seen at the corners in the range $(-\pi/4\leq k_{x}\leq \pi/4, -\pi/4\leq k_{y}\leq \pi/4)$. Same calculation for the DW phase, in a symmetric gauge potential, shows the existence small peaks at the BZ (doubly reduced) corners, in addition to the peak at $(0,0)$, as shown in fig (\ref{symmcomp}(b)).\\

\begin{figure}[ht]
\centerline{ \epsfxsize 12cm \epsfysize 7cm \epsffile{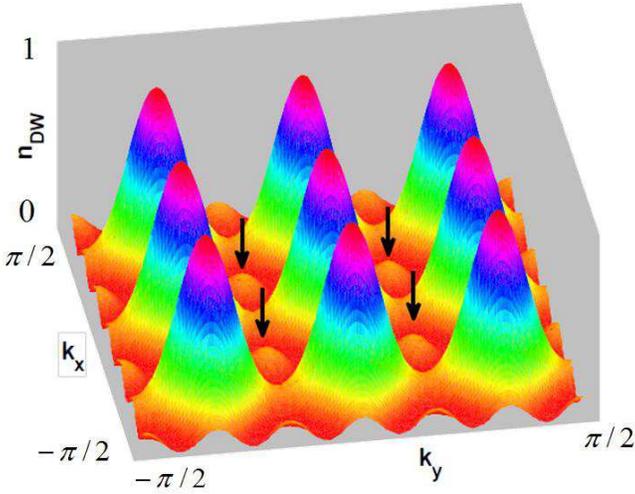}}
\caption{\textit{(Color Online)} Extra small peaks in the DW momentum distribution $n(\bs{k})$ symmetric gauge potential. The arrows show the location of the extra peaks at intermediate points}
\label{dwpeaks}
\end{figure}

The effect of application of symmetric gauge potential to the system is equivalent to rotating the system \cite{goldbaum, rashi}
and thus, leads to formation of vortices
in the system when the superfluid order parameter is finite. Since we perform our calculations at the phase boundary, the results for the momentum distribution of the DW and MI phases
in presence of symmetric gauge potential, provides information about
the structure of vortex in a supersolid (at the  DW-SS  phase boundary ) and vortex in a superfluid
(at the MI-SF phase boundary).
The preceeding discussion demonstrates that rotating supersolid reflects some extra peaks in the momentum distribution in addition to the peaks observed in a rotating ordinary superfluid. These extra peaks are clearly demonstrated in fig (\ref{dwpeaks}) and is  one of  the central result of this work.\\
Since momentum distribution can be measured using  TOF absorption imaging technique, this  provides a way to experimentally distinguish  between the supersolid phase  and superfluid phase by comparing  the respective vortex profile. To analyze this issue further in the next section we compare the momentum distribution
$n (\bs{k})$ in symmetric gauge to the momentum distribution corresponding to the many body states having definite quasi-angular momentum in a symmetric gauge potential. This quasi-angular momentum are analogues of the Bloch-momentum, but for a rotationally invariant system and has been discussed in detail \cite{bhat}

\subsection{Quasi-angular momentum distribution}
In this section we re evaluate the momentum distribution $n (\bs{k})$ but for a fixed quasi angular momentum and
compare this with $n(\bs{k})$  demonstrated in Fig. \ref{symmcomp}.

 To  simplify calculation we approximate the wannier functions as  delta functions to  calculate the
 correlation function with a fixed value of angular momentum given to the system. The field operators
can now be expanded as
\beq\psi(\mathbf{r})=\frac{1}{\sqrt{M}}\sum_{l}e^{i2\pi ml/M}\delta(\mathbf{r}-\mathbf{r}_{l}) b_{l}\eeq
where $m$ is the quasi-angular momentum, $M$ is the number of lattice sites.  Substituting this in the expression
(\ref{mdf}) yields
 \bea n(\mathbf{k})=\frac{1}{M}\sum_{l,l^{'}}e^{i2\pi m(l-l^{'})/M}\delta^{*}(\mathbf{r}-\mathbf{r}_{l})\delta(\mathbf{r}-\mathbf{r}_{l^{'}})
 \nonumber\\ \times<\psi_{ins}|b_{l}^{\dag}b_{l^{'}}|\psi_{ins}>e^{i\mathbf{k}.
(\mathbf{r}_{l}-\mathbf{r_{l^{'}}})}\label{momqu}\eea
 The summation indices $l\epsilon\left\{A,B\right\}$ and $l^{'}\epsilon\left\{A,B\right\}$ includes the entire lattice.\\
 As compared to  the expresssion (\ref{nk}) , in (\ref{momqu})  the system has a prefixed value of quasi angular momentum  and therefore
we can associate a vortex phase with the system.

\begin{figure}
\begin{center}
\centerline{\epsfxsize 5cm \epsffile{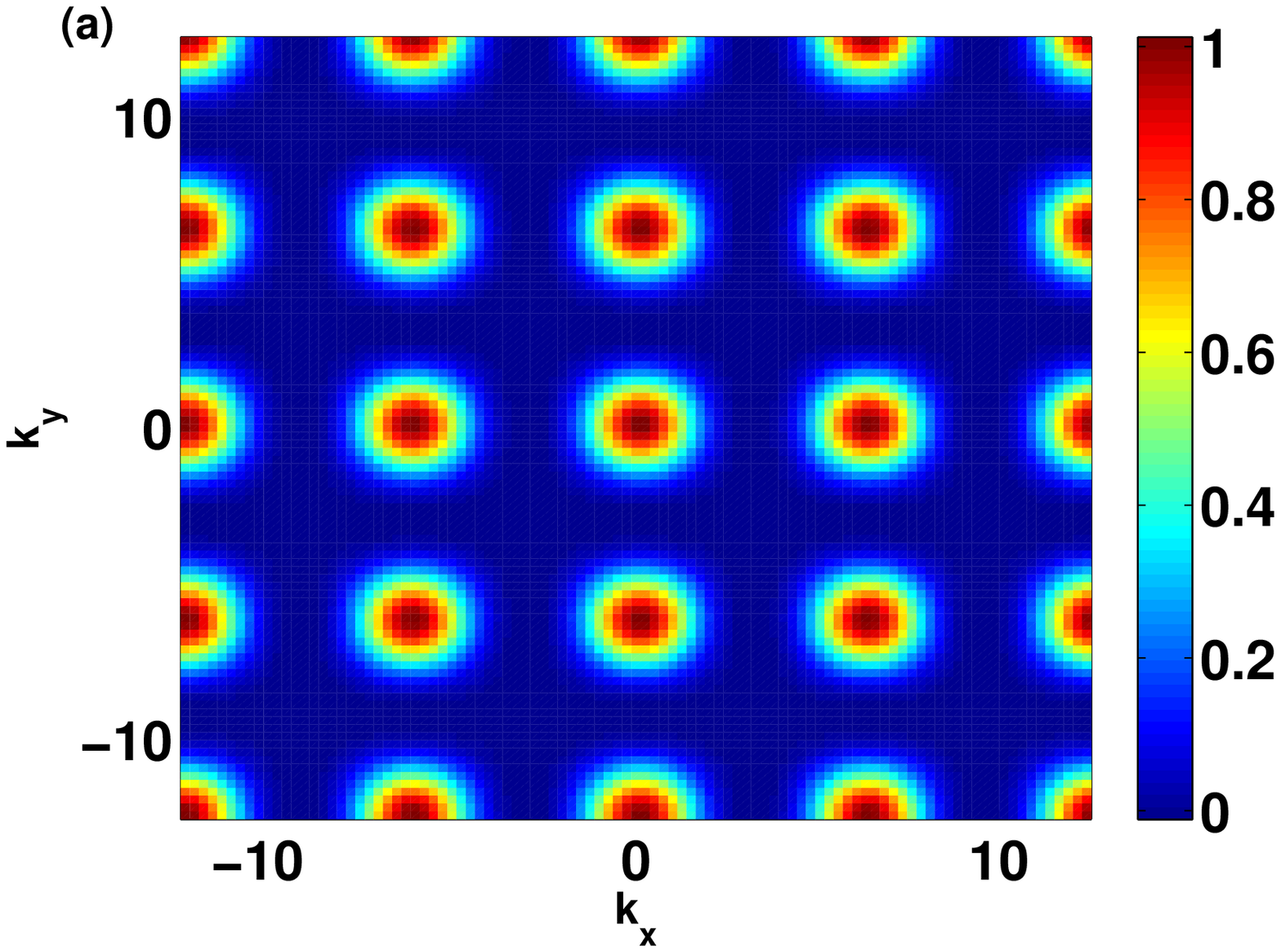}\epsfxsize 5cm \epsffile{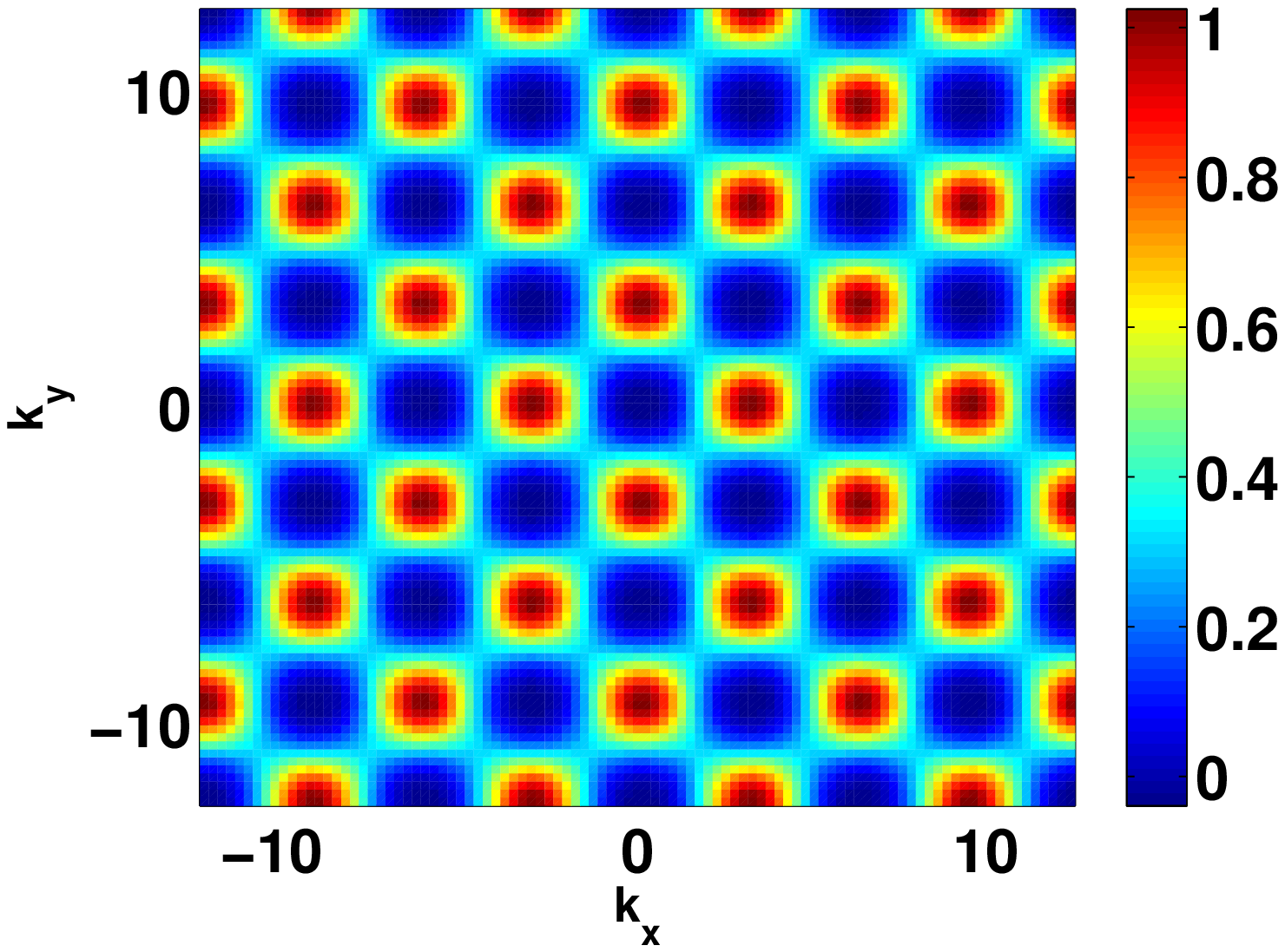}}
\caption{\textit{(Color Online)} Quasi angular momentum distribution of the DW phase for (a) m=0 ,(b) m=1, plotted over a range
$-4\pi\leq k_{x},k_{y} \leq 4\pi$}
\label{quasidw}
\end{center}
\end{figure}

\begin{figure}[ht]
\centerline{ \epsfxsize 8.5cm \epsfysize 6cm \epsffile{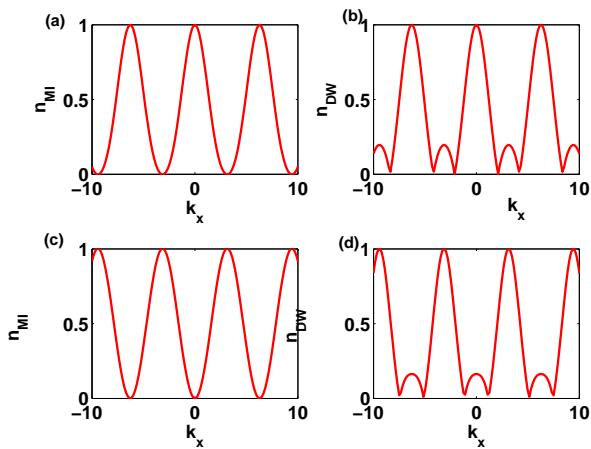}}
\caption{\textit{(Color Online)} The cross sectional plots for m=0 and m=1 quasi angular momentum state for MI((a),(c)) and DW ((b), (d)) phases.The DW phase shows distinctive peaks compared to MI phase}
\label{quasimidw}
 \end{figure}

For DW and MI phase, the quasi angular momentum distribution is given by :
\begin{widetext}
\bea n_{DW}(\mathbf{k})=\frac{1}{M}\sum_{l,l^{'}}e^{i2\pi m(l-l^{'})/M}\delta^{*}(\mathbf{r}-\mathbf{r}_{l})\delta(\mathbf{r}-\mathbf{r}_{l^{'}})e^{i\mathbf{k}.
(\mathbf{r}_{l}-\mathbf{r}_{l^{'}})} (\frac{2n_{0}-1}{2}
-\left[\frac{n_{0}^{2}}{V(z-1)}
+\frac{(n_{0}^{2}-1)}{2U-V(z+1)}\right]t\nonumber\\
+2n_{0}\left[\frac{n_{0}^{2}}{2V^{2}(z-1)^{2}}+\frac{(n_{0}^{2}-1)}{2[2U-V(z+1)]^{2}}
+\frac{n_{0}^{2}}{UV(z-1)}+\frac{(n_{0}^{2}-1)}{U[2U-V(z+1)]}\right]t^{2}+O(t^{3}))\nonumber \eea

\end{widetext}
Similar expression for $n(\bs{k})$  with a given quasi angular momentum for the MI phase can also
 be evaluated withing strong coupling expansion. For a square lattice with four sites, Fig. \ref{quasidw} (a) shows the momentum distribution for $m=0$ for DW phase. The $m=0$ state has a peak at $k=2\pi n$, with $n$ as an integer. However for
higher quasi angular momentum state m=1, $n( \bs{k})$ vanishes at $k = 2 \pi n$ in Fig. \ref{quasidw} (b) . Again the DW-SS phase boundary can
be separated from the MI-SF phase boundary  by noting the appearance of small extra peaks in the former case. To demonstrate these small peaks clearly we plot the  cross sectional plots for the DW phase in  Fig. \ref{quasimidw}.

 We also plot the quasi momentum distribution for the state $m=1$ for MI and DW phases.

Quasi angular momentum is connected with the phase information and hence, the vorticity. The fact that  a zero quasi angular momentum state can be distinguished from a non-zero one by looking at the corresponding $n (\bs{k})$, allows the experimenters to verify that vorticity has entered in the system through the TOF measurement.



\section{summary and conclusion}
To summarize we have determined the modification of the DW-SS and MI-SF phase boundary corresponding to EBHM at zero temperature due
to the effect of an artificial gauge field within the framework of strong coupling perturbation theory. Using the same approach we calculated the
momentum distribution at this phase boundary which can be verified experimentally using the time of flight imaging. The momentum distribution
reflects the symmetry of the gauge potential. The momentum distribution shows distinctive feature at the DW-SS phase boundary as compared to the MI-SF phase boundary. Particulaly evaluating the momentum distribution for states with definite quasi-angular momentum  at such phase boundary in a symmetric gauge we clearly demonstrate how this can be used to probe the vortex profile of the SS
phase at the DW-SS boundary due to the action of the gauge field. There have been significant progress in cold atom experiments in identifying the enigmatic supersolid phase in ultra cold atomic systems \cite{expt1, expt2, lev}. Our calculation will  hopefully augment further study in the behavior of such SS phase
in presence of an artificial gauge field.

RS is supported by a fellowship given by CSIR and the work of SG is supported by a grant from Planning section, IIT Delhi.

\appendix

\include{Appendix}
\section{Perturbation corrections to the  ground state and particle hole excitation energy in
the DW and MI phases in presence of artificial magnetic field}
\subsection{Corrections to grounds state energy}
Here we provide the detail expressions of $E_{n}^{(0)}, E_{n}^{(1)}, E_{n}^{(2)}, E_{n}^{(3)}$ that appears in the equation
(\ref{enins}) for DW phase. The zeroth order term is given by
\bea E_{DW}^{(0)} & = & \langle \Psi_{DW}^{(0)}|H_{0}|\Psi_{DW}^{(0)}\rangle \nonumber\\
& = & M(\frac{n_{A}(n_{A}-1)+n_{B}(n_{B}-1)}{4}+zV\frac{n_{A}n_{B}}{2} \nonumber \\
&  & \mbox{} -\mu\frac{n_{A}+n_{B}}{2}) \label{dwzero}\eea
The first order correction is given as
\bea E_{DW}^{(1)} & = &  \langle \Psi_{DW}^{(0)}|H_{P}|\Psi_{DW}^{(0)} \rangle=  0\label{dwone}\eea
This is because $\langle n_{A}-1|n_{A}\rangle_{i}=\langle n_{B}-1|n_{B}\rangle_{j}=0$, where $i$ and $j$ can be any site. For the same reason all odd order correction like $E_{n}^{(3)}$  also vanishes.
The second order correction, which is also the first non-vanishing correction is given by
\begin{widetext}
\bea E_{DW}^{(2)} & = & \sum_{m\neq n}\frac{| \langle \Psi_{DW}(m_{A},m_{B})|H_{P}|\Psi_{DW}^{(0)}(n_{A},n_{B}) \rangle |^{2}}{(E_{n_{A,B}}-E_{m_{A,B}})}   \nonumber \\
 & = & \left[\frac{n_{A}(n_{B}+1)}{(n_{A}-n_{B}-1)+(zn_{B}-zn_{A}+1)V}+\frac{n_{B}(n_{A}+1)}{(n_{B}-n_{A}-1)+(zn_{A}-zn_{B}+1)V
}\right]Mzt^{2}
\label{en2dw}\eea
\end{widetext}
Substituting the above expressions in the right hand side of Eq. (\ref{enins}) one gets the ground state energy of the DW
state upto the third order as
\begin{widetext}
\bea \frac{E_{DW}(n_{A},n_{B})}{M} & = &\frac{n_{A}(n_{A}-1)+n_{B}(n_{B}-1)}{4}+zV\frac{n_{A}n_{B}}{2}-\mu\frac{n_{A}+n_{B}}{2}\nonumber\\
& & \mbox{}+\left[\frac{n_{A}(n_{B}+1)}{(n_{A}-n_{B}-1)+(zn_{B}-zn_{A}+1)V}+\frac{n_{B}(n_{A}+1)}{(n_{B}-n_{A}-1)+(zn_{A}-zn_{B}+1)V
}\right]zt^{2}+O(t^{4})\nonumber\\
& & \label{dwgdenergy}\eea
\end{widetext}
Setting $n_A=n_0$ and $n_B=n_0-1$ in the above expression we get the expression (\ref{endwgd}).

Also, we can get the corresponding corrections for the ground state in MI phase  by substituting in the above expressions for the DW state $n_{A}=n_{B}=n_0$. The ground state energy in the MI Phase given in Eq. (\ref{enmigd}) can be obatined from the expression (\ref{dwgdenergy}) by putting $n_{A}=n_{B}=n_{0}$.
\subsection{Calculation of energy of DW state with an extra particle or hole}
 Explicitly written wave function of the DW phase with an extra particle is
 \bea |\Psi_{DW}^{par(0)}> & = & \sum_{j\epsilon B, j=1}^{M/2}f_{j}^{DW(B)}|n_{A}^{(1)},n_{A}^{(3)} ... n_{A}^{(M/2)};\nonumber\\
& & n_{B}^{(2)} ....  n_{B}^{(j-1)},n_{B}^{(j)}+1 ... n_{B}^{(M)}\rangle\nonumber\eea
where $f_{j}^{DW(B)}$ is the eigenvector of $\sum_{i}t_{ji}t_{ij^{'}}$ with the minimum eigenvalue ($\epsilon^{2}t^{2}$)$f_{j}^{DWB}$, such that $\sum_{i,j^{'}}t_{ji}t_{ij^{'}}f_{j^{'}}^{DW(B)}=\epsilon^{2}t^{2}f_{j}^{DW(B)}$,

The zeroth order energy of the DW phase is
\bea E_{DW}^{par(0)} & = & <\Psi_{DW}^{par(0)}|H_{0}|\Psi_{DW}^{par(0)}>\nonumber\\
& = &  E_{DW}^{(0)}+n_{B}+zVn_{A}-\mu\label{dwpar0}\eea
First order correction in energy
\bea E_{DW}^{par(1)}& = & <\Psi_{DW}^{par(0)}|H_{P}|\Psi_{DW}^{par(0)}>\nonumber\\
& = & 0\label{dwpar1}\eea
Again, all odd order terms vanishes. The second order correction to the energy of DW phase with an extra particle is given as
\begin{widetext}
\bea E_{DW}^{par(2)} & = & \sum_{m\neq n}\frac{|<\Psi_{DW}^{par}(m_{A},m_{B})|H_{P}|\Psi_{DW}^{par(0)}(n_{A},n_{B})>|^{2}}{(E_{n_{A,B}}-E_{m_{A,B}})}\nonumber\\
& = & E_{DW}^{(1)}+\left[\frac{(n_{A}+1)(n_{B}+1)\epsilon^{2}t^{2}}{(n_{B}-n_{A})+(zn_{A}-zn_{B})V}-
\frac{n_{A}(n_{B}+1)\epsilon^{2}t^{2}}{(n_{A}-n_{B}-1)+(zn_{B}-zn_{A}+1)V}\right]\nonumber\\
& & \mbox{} +\left[-\frac{n_{B}(n_{A}+1)\epsilon^{2}t^{2}}{(n_{B}-n_{A}-1)+(zn_{A}-zn_{B}+1)V}+\frac{n_{A}(n_{B}+2)zt^{2}}
{(n_{A}-n_{B}-2)+(zn_{B}-zn_{A}+2)V}\right]\nonumber\\
& & \mbox{} +\left[\frac{n_{B}(n_{A}+1)(\epsilon^{2}-z)t^{2}}{(n_{B}-n_{A}-1)+(zn_{A}-zn_{B})V}+
\frac{2n_{A}(n_{B}+1)(\epsilon^{2}-z)t^{2}}{(n_{A}-n_{B}-1)+(zn_{B}-zn_{A}+2)V}\right]\label{dwpar2}\eea
\end{widetext}


Thus, adding equations (\ref{dwpar0}), (\ref{dwpar2})  gives us the energy of DW state with an extra particle upto
the third order.

In the same way, the energy of the DW phase with an extra hole can be calculated and is given as
\begin{widetext}
\bea E_{DW}^{hol}(n_{A},n_{B}) & = & E_{DW}^{ins}(n_{A},n_{B})-(n_{A}-1)-zVn_{B}+\mu+\nonumber\\
& & \mbox{} +\left[\frac{n_{A}n_{B}\epsilon^{2}t^{2}}{(n_{B}-n_{A})+(zn_{A}-zn_{B})V}-
\frac{n_{A}(n_{B}+1)\epsilon^{2}t^{2}}{(n_{A}-n_{B}-1)+(zn_{B}-zn_{A}+1)V}\right]\nonumber\\
& & \mbox{} +\left[-\frac{n_{B}(n_{A}+1)\epsilon^{2}t^{2}}{(n_{B}-n_{A}-1)+(zn_{A}-zn_{B}+1)V}+\frac{(n_{A}-1)(n_{B}+1)zt^{2}}
{(n_{A}-n_{B}-2)+(zn_{B}-zn_{A}+2)V}\right]\nonumber\\
& & \mbox{} +\left[\frac{n_{B}(n_{A}+1)(\epsilon^{2}-z)t^{2}}{(n_{B}-n_{A}-1)+(zn_{A}-zn_{B})V}+
\frac{2n_{A}(n_{B}+1)(\epsilon^{2}-z)t^{2}}{(n_{A}-n_{B}-1)+(zn_{B}-zn_{A}+2)V}\right]+O(t^{4})\eea
\end{widetext}
Again  the expressions (\ref{endwpar})  and (\ref{endwhol}) are  obtained by substituting $n_{A}=n_{0}$ and $n_{B}= n_{0}-1$.

\subsection{Calculation of energy of MI state with an extra particle or hole}
The MI phase with an extra particle has the following wave function,
\bea |\Psi_{MI}^{par(0)}> & = & \sum_{j=1}^{M}f_{j}^{MI}|n_{0}^{(1)},n_{0}^{(2)} ... n_{0}^{(j)}+1, ... n_{0}^{(M)}\rangle\nonumber\eea
where  the coefficient $f_{j}$ are to be determined  from the lowest eigenvalue $\sum_{j^{'}}t_{jj^{'}}f_{j^{'}}^{MI}=ztf_{j}^ {MI}$ as explained earlier.
The zeroth order correction to the energy of MI phase with an extra particle is now given as
\bea E_{MI}^{par(0)} & = & <\Psi_{MI}^{par(0)}|H_{0}|\Psi_{MI}^{par(0)}>\nonumber\\
& = &  E_{MI}^{(0)}+n_{0}+zVn_{0}-\mu\label{mipar0}\eea
The first order correction to the energy of this state is
\beq E_{MI}^{par(1)}=<\Psi_{MI}^{par(0)}|H_{P}|\Psi_{MI}^{par(0)}>=-(n_{0}+1)\epsilon t  \label{mipar1} \eeq
The second order energy correction for the extra particle MI phase is as calculated below
\bea E_{MI}^{par(2)}& = & \sum_{m\neq n}\frac{|<\Psi_{MI}^{par}(m_{0})|H_{P}|\Psi_{MI}^{par(0)}(n_{0})>|^{2}}{(E_{n}^{0}-E_{m}^{0})}\nonumber\\
& = &E_{MI}^{(2)} \nonumber \\
&  & \mbox{} + n_{0}(n_{0}+1)\left[(z-\epsilon^{2})+\frac{2(z-\epsilon^{2})}{(1-2V)}+\frac{2\epsilon^{2}}{(1-V)}\right]t^2 \nonumber\\
& & \mbox{}-\frac{n_{0}(n_{0}+2)z}{2(1-V)}t^{2} \label{mipar2} \eea

\begin{widetext}
\bea E_{MI}^{par(3)} & = & \sum_{k\neq n}\sum_{m\neq n}\frac{<\Psi_{MI}^{par(0)}(n_{0})|H_{P}|\Psi_{MI}^{par}(m_{0})><\Psi_{MI}^{par}(m_{0})|H_{P}|\Psi_{MI}^{par}(k_{0})><\Psi_{MI}^{par}(k_{0})|H_{P}|\Psi_{MI}^{par}(n_{0})>}
{(E_{m}^{0}-E_{n}^{0})(E_{k}^{0}-E_{m}^{0})}\nonumber\\
& & \mbox{} -<\Psi_{MI}^{par(0)}(n_{0})|H_{P}|\Psi_{MI}^{par(0)}(n_{0})>\sum_{m\neq n}\frac{|<\Psi_{MI}^{par(0)}(n_{0})|H_{P}|\Psi_{MI}^{par}(m_{0})>|^{2}}{(E_{m}^{0}-E_{n}^{0})^{2}}\nonumber\\
& = & -n_{0}(n_{0}+1)\left\{n_{0}\left[(z-2\epsilon)\epsilon+\frac{(\epsilon^{2}-3z+3)\epsilon}{(1-V)^{2}}\right]\right\}t^{3}\nonumber\\
& & \mbox{} -n_{0}(n_{0}+1)\left\{(n_{0}+1)\left[(z-\epsilon^{2})\epsilon-\frac{(2\epsilon^2-6z+6)\epsilon}{(1-V)^{2}}+
\frac{2\epsilon(z-\epsilon^{2})}{(1-2V)^{2}} +\frac{2\epsilon(\epsilon^{2}-3z+3)}{(1-V)}+\frac{4(z-2)\epsilon}{(1-2V)}\right]\right\}t^{3}\nonumber\\
& & \mbox{}-n_{0}(n_{0}+1)(n_{0}+1)\frac{4\epsilon(\epsilon^{2}-3z+3)}{(1-V)(1-2V)}t^{3}
 -n_{0}(n_{0}+1)(n_{0}+2)
 \left[\frac{\epsilon(z-1)}{(1-V)}
-\frac{z\epsilon}{4(1-V)^{2}}\right]t^{3}\label{mipar3}\eea
\end{widetext}
Adding Eqs. (\ref{mipar0}), (\ref{mipar1}), (\ref{mipar2}) and (\ref{mipar3}) gives the energy expression for MI phase with an extra particle given in  (\ref{enmipar}). This expression recovers the known result of onsite Bose Hubbard model when $V=0$ \cite{Freericks}.

The energy for the Mott phase with an extra hole is calculated, in the same way as above and is given in eq (\ref{enmihol}).

\label{app1}
\section{Perturbative correction to the wave function and momentum distribution} 
\subsection{Corrections to the wave function for the insulating DW phase}
We start with the definition of the ground state wave function for the DW phase eq(\ref{dwwfgd})( $2$x$2$ lattice, for simplicity).
We apply perturbation theory on $|\Psi_{DW}^{(0)}\rangle$ to calculate $|\Psi_{DW}^{ins}\rangle$ up to the desired order eq(\ref{genwf}).\\
Using the diagrams given in  Fig. (\ref{1stexcited}), the $1$st order correction to the ground state 
wave function  of DW state in for a unit cell in a square lattice 
(for all possible $|m\rangle$'s) can be explicitly written as 
\begin{widetext}
\bea |\Psi_{DW}^{(1)}\rangle & = & \sum_{m\neq|\Psi_{DW}^{(0)}\rangle}\frac{-\sum_{ll^{'}}t_{ll^{'}} \langle m|b_{l}^{\dag}b_{l^{'}}|0\rangle}{E_{0m}}|m\rangle\nonumber\\
  & = & \frac{-\sqrt{n_{B}(n_{A}+1)}}{E_{1}}[t_{12}|n_{A}^{(1)}+1,n_{B}^{(2)}-1,n_{A}^{(3)},n_{B}^{(4)}\rangle+t_{14}|n_{A}^{(1)}+1,n_{B}^{(2)},
  n_{A}^{(3)},n_{B}^{(4)}-1\rangle\nonumber\\
  & & \mbox{} +t_{32}|n_{A}^{(1)},n_{B}^{(2)}-1,n_{A}^{(3)}+1,n_{B}^{(4)}\rangle+t_{34}|n_{A}^{(1)}, n_{B}^{(2)}, n_{A}^{(3)}+1,n_{B}^{(4)}-1\rangle]\nonumber\\
& & \mbox{}-\frac{\sqrt{n_{A}(n_{B}+1)}}{E_{2}}[t_{21}|n_{A}^{(1)}-1, n_{B}^{(2)}+1, n_{A}^{(3)}, n_{B}^{(4)}\rangle+t_{23}|n_{A}^{(1)}, n_{B}^{(2)}+1, n_{A}^{(3)}-1, n_{B}^{(4)}\rangle\nonumber\\
& & \mbox{}+t_{41}|n_{A}^{(1)}-1, n_{B}^{(2)}, n_{A}^{(3)}, n_{B}^{(4)}+1\rangle+t_{43}|n_{A}^{(1)}, n_{B}^{(2)}, n_{A}^{(3)}-1, n_{B}^{(4)}+1\rangle]\nonumber\\
& & \label{wf1}\eea
where
\bea
 E_{1} & = & (n_{B}-n_{A}-1)+(zn_{A}-zn_{B}+1)V\nonumber\\
 E_{2} & = & (n_{A}-n_{B}-1)+(zn_{B}-zn_{A}+1)V\nonumber\eea
\end{widetext}

To calculate the second order correction  using eq(\ref{2ndint}) following Fig. (\ref{2ndexcited})
we consider  $|m\rangle=|n_{A}^{(1)}+1, n_{B}^{(2)}+1, n_{A}^{(3)}, n_{B}^{(4)}\rangle$ 
\beq \frac{\langle m|H_{P}|0\rangle
}{E_{0m}}  =  \frac{-\sqrt{n_{B}(n_{A}+1)}t_{12}}{E_{1}}\label{m} \eeq
 The corresponding possibilities for $|m^{'}\rangle$ for $|m\rangle=|n_{A}^{(1)}+1, n_{B}^{(2)}-1, n_{A}^{(3)}, n_{B}^{(4)}\rangle$  are represented in Fig. (\ref{2ndexcited}(c)-(i)). For illustration let us choose 
$|m^{'}\rangle=|n_{A}^{(1)}+2, n_{B}^{(2)}-2, n_{A}^{(3)}, n_{B}^{(4)}\rangle$. Then 
\bea E_{0m^{'}} & = & E_{0}-E_{m^{'}}=E_{01}\nonumber\\
& = & 2(n_{A}-n_{B}-2)+(2zn_{A}-2zn_{B}+4)V\nonumber\eea
 The contribution of this $|m^{'}\rangle$ state to the second order correction to DW wave function is
 \bea |\Psi_{DW}^{(2)}\rangle_{a_{1}} & = & \frac{\sqrt{n_{B}(n_{A}+1)(n_{A}+2)(n_{B}-1)}}{(E_{1})(E_{01})}t_{12}^{2}\nonumber\\
& &  |n_{A}^{(1)}+2, n_{B}^{(2)}-2, n_{A}^{(3)}, n_{B}^{(4)}\rangle\nonumber\eea
 Similarly, we calculate the contribution from all of the $|m^{'}\rangle$ states corresponding to this particular $|m\rangle$ .
Summing over this contributions we 
 the second order correction to the wave function due to this $| m \rangle$ state which looks like
\begin{widetext}
\bea |\Psi_{DW}^{(2)}( | m \rangle )\rangle_{a} & = & \frac{\sqrt{n_{B}(n_{A}+1)(n_{A}+2)(n_{B}-1)}}{(E_{1})(E_{01})}t_{12}^{2}|n_{A}^{(1)}+2, n_{B}^{(2)}-2, n_{A}^{(3)}, n_{B}^{(4)}\rangle \nonumber\\
&  & \mbox{} +\frac{n_{B}\sqrt{(n_{A}+1)(n_{A}+2)}}{(E_{1})(E_{02})}t_{12}t_{14}|n_{A}^{(1)}+2, n_{B}^{(2)}-1, n_{A}^{(3)}, n_{B}^{(4)}-1\rangle\nonumber\\
& & \mbox{} +\frac{(n_{A}+1)\sqrt{n_{B}(n_{B}+1)}}{(E_{1})(E_{03})}t_{12}t_{41}|n_{A}^{(1)}, n_{B}^{(2)}-1, n_{A}^{(3)}, n_{B}^{(4)}+1\rangle\nonumber\\
 &  & \mbox{}+\frac{(n_{A}+1)\sqrt{n_{B}(n_{B}-1)}}{(E_{1})(E_{04})}t_{12}t_{32}|n_{A}^{(1)}+1, n_{B}^{(2)}-2, n_{A}^{(3)}+1, n_{B}^{(4)}\rangle\nonumber\\
& & \mbox{} +\frac{n_{B}\sqrt{n_{A}(n_{A}+1)}}{(E_{1})(E_{05})}t_{12}t_{23}|n_{A}^{(1)}+1, n_{B}^{(2)}, n_{A}^{(3)}-1, n_{B}^{(4)}\rangle\nonumber\\
 &  & \mbox{}+\frac{\sqrt{n_{A}n_{B}(n_{A}+1)(n_{B}+1)}}{(E_{1})(E_{06})}t_{12}t_{43}|n_{A}^{(1)}+1, n_{B}^{(2)}-1, n_{A}^{(3)}-1, n_{B}^{(4)}+1\rangle\nonumber\\
& & \mbox{} +\frac{n_{B}(n_{A}+1)}{(E_{1})(E_{07})}t_{12}t_{34}|n_{A}^{(1)}+1, n_{B}^{(2)}-1, n_{A}^{(3)}, n_{B}^{(4)}-1\rangle\nonumber\eea
\end{widetext}
Follwoing the same procedure for all other $| m \rangle$ states defined in Fig. (\ref{1stexcited}) and 
 adding all contributions, we get the total second order correction to wave-function for the DW phase.

\subsection{ Terms in momentum distribution calculation}
The  moemntum distribution in  (\ref{nk}) can be written as 
\bea n(\mathbf{k})& = &\frac{|W(\mathbf{k})|^{2}}{M}\sum_{l,l^{'}}\langle \Psi_{DW}|b_{l}^{\dag}b_{l^{'}}|\Psi_{DW}\rangle e^{-i\mathbf{k}.
(\mathbf{R}_{l}-\mathbf{R}_{l^{'}})}\nonumber\\
& = & n(\mathbf{k})^{(0)}+n(\mathbf{k})^{(1)}+n(\mathbf{k})^{(2)}\nonumber\eea
where
\begin{widetext}
\bea n(\mathbf{k})^{(0)} & = & \frac{|W(\mathbf{k})|^{2}}{M}\sum_{l,l^{'}}\{\langle \Psi_{DW}^{(0)}|b_{l}^{\dag}b_{l^{'}}|\Psi_{DW}^{(0)}\rangle\}e^{-i\mathbf{k}.(\mathbf{R}_{l}
-\mathbf{R}_{l^{'}})}\nonumber\\
& & \label{nk0}\\
n(\mathbf{k})^{(1)} & = & \frac{|W(\mathbf{k})|^{2}}{M}\sum_{l,l^{'}} (\langle\Psi_{DW}^{(0)}|b_{l}^{\dag}b_{l^{'}}|\Psi_{DW}^{(1)}\rangle\nonumber\\
& & \mbox{}+\langle\Psi_{DW}^{(1)}|b_{l}^{\dag}b_{l^{'}}|\Psi_{DW}^{(0)}\rangle)e^{-i\mathbf{k}.(\mathbf{R}_{l}
-\mathbf{R}_{l^{'}})}\nonumber\\
& & \label{nk1}\\
n(\mathbf{k})^{(2)} & = & \frac{|W(\mathbf{k})|^{2}}{M}\sum_{l,l^{'}}(\langle\Psi_{DW}^{(0)}|b_{l}^{\dag}b_{l^{'}}|\Psi_{DW}^{(2)}\rangle\nonumber\\
& & \mbox{} +\langle\Psi_{DW}^{(1)}|b_{l}^{\dag}b_{l^{'}}|\Psi_{DW}^{(1)}\rangle
 +\langle\Psi_{DW}^{(2)}|b_{l}^{\dag}b_{l^{'}}|\Psi_{DW}^{(0)}\rangle)\nonumber\\
& &  e^{-i\mathbf{k}.(\mathbf{R}_{l}
-\mathbf{R}_{l^{'}})}\nonumber\\
& & \label{nk2}\eea
\end{widetext}
Here we keep terms only upto second order and   terms $\langle\Psi_{DW}^{(1)}|b_{l}^{\dag}b_{l^{'}}|\Psi_{DW}^{(2)}\rangle $, $\langle\Psi_{DW}^{(2)}|b_{l}^{\dag}b_{l^{'}}|\Psi_{DW}^{(1)}\rangle$ and $\langle\Psi_{DW}^{(0)}|b_{l}^{\dag}b_{l^{'}}|\Psi_{DW}^{(2)}\rangle$ are neglected. 
Using the wave function from eq(\ref{dwwfgd}) we  get  g
\begin{widetext}
\bea n(\mathbf{k})^{(0)} & = & |W(\mathbf{k})|^{2}\left(\frac{n_{A}+n_{B}}{2}\right)\label{mom0}\eea
\bea n(\mathbf{k})^{(1)} & = & -\frac{n_{B}(n_{A}+1)}{E_{1}}\frac{|W(\mathbf{k})|^{2}}{M/2} \sum_{l,l^{'}}
t_{ll^{'}}e^{i\mathbf{k}.(\mathbf{R}_{l}
-\mathbf{R}_{l^{'}})}  -\frac{n_{A}(n_{B}+1)}{E_{2}}\frac{|W(\mathbf{k})|^{2}}{M/2} \sum_{l,l^{'}}
t_{ll^{'}}e^{i\mathbf{k}.(\mathbf{R}_{l}
-\mathbf{R}_{l^{'}})}\nonumber\\
& = & |W(\mathbf{k})|^{2}\left[\frac{n_{B}(n_{A}+1)}{E_{1}}+\frac{n_{A}(n_{B}+1)}{E_{2}}
\right]\epsilon(\mathbf{k}) \label{mom1}\eea
and 
\bea n(\mathbf{k})^{(2)} & = & \frac{|W(\mathbf{k})|^{2}}{M/2}\left[\frac{n_{B}(n_{A}+1)}{2E_{1}^{2}}+\frac{n_{A}(n_{B}+1)}{2E_{2}^{2}}-
\frac{n_{B}(n_{A}+1)}{E_{1}}-\frac{n_{A}(n_{B}+1)}{E_{2}}\right]
\times (n_{B}+n_{A}+1)
[\sum_{l,l^{'},l^{''}}t_{ll^{'}}t_{l^{'}l^{''}}e^{i\mathbf{k}.(\mathbf{R}_{l}-
\mathbf{R}_{l^{''}})}]\nonumber\\
& = &|W(\mathbf{k})|^{2}[\frac{n_{B}(n_{A}+1)}{2E_{1}^{2}}+\frac{n_{A}(n_{B}+1)}{2E_{2}^{2}}-
\frac{n_{B}(n_{A}+1)}{E_{1}}-\frac{n_{A}(n_{B}+1)}{E_{2}}]
\times (n_{B}+n_{A}+1)(\epsilon^{2}(\mathbf{k})-2dt^{2})\label{mom2}\eea
\end{widetext}

\label{app2}

\end{document}